\documentclass[12pt]{article}
\input epsf.sty
\usepackage{mathrsfs}
\usepackage{amsmath}
\usepackage{mathrsfs}
\usepackage{amssymb}
\usepackage{color}
\usepackage{psfrag}
\usepackage{graphicx}
\usepackage{subfigure}

\usepackage{rotating}

\newcommand{\bea}{\begin{eqnarray}}
\newcommand{\eea}{\end{eqnarray}}
\newcommand{\be}{\begin{equation}}
\newcommand{\ee}{\end{equation}}
\newcommand{\vs}[1]{\vspace{#1 mm}}

\newcommand{\dsl}{\pa \kern-0.5em /}

\newcommand{\half}{\frac{1}{2}}
\newcommand{\pa}{\partial}

\newcommand{\nn}{\nonumber\\}

\begin{document}
\topmargin 0pt
\oddsidemargin 0mm

\begin{flushright}


\end{flushright}

\vspace{2mm}

\begin{center}
{\Large \bf From AdS to Schr\" odinger/Lifshitz dual space-times without or with hyperscaling violation}  

\vs{10}

{Parijat Dey\footnote{E-mail: parijat.dey@saha.ac.in} and 
Shibaji Roy\footnote{E-mail: shibaji.roy@saha.ac.in}}

 \vspace{4mm}

{\em

 Saha Institute of Nuclear Physics,
 1/AF Bidhannagar, Calcutta-700 064, India\\}

\end{center}

\vs{10}

\begin{abstract}
It is observed that the (intersecting) branes of M/string theory, which are known
to give AdS geometry (directly or upto a conformal transformation) in the near horizon limit,
do also lead to Schr\" odinger/Lifshitz dual space-times (without or with 
hyperscaling violation) upon
using appropriate solution generating transformation and dimensional reduction.
We show that the dynamical exponents of the Schr\" odinger and the Lifshitz space-times
obtained in this way always add upto 2. We illustrate this by several
examples, including M2-, M5-branes of M-theory and D$(p+1)$-branes ($p\neq 4$,
since in this case the near horizon limit does not give AdS geometry) of string theory
as well as many of their intersecting solutions. The Schr\" odinger space-time can be 
obtained by the standard wave generating technique along one of the brane
directions (for single brane) or one of the common brane directions (for intersecting branes)
and then interchanging the light-cone coordinates by double Wick rotations,
whereas, the Lifshitz space-time can be obtained by dimensionally reducing 
(for M-theory) along the
wave direction or taking T-duality (for string theory) along the same direction. 
We thus obtain Schr\" odinger/Lifshitz dual space-times without or with hyperscaling
violation from the same M/string theory solutions and they preserve some fraction of 
the supersymmetry.      
\end{abstract}

\newpage
\section{Introduction}

Both Schr\" odinger and Lifshitz symmetries are non-relativistic symmetries as the 
space and the time scale differently: $t \to \lambda^z t$, $x^i \to \lambda x^i$, 
($z \neq 1$), where $i=1,2,\ldots,d$, $d$ being the number of spatial dimensions and 
$z$, the dynamical critical exponent, unlike the relativistic symmetry 
where the space-time scale as $t \to \lambda t$, $x^i \to \lambda x^i$. However, the 
Schr\" odinger symmetry is much larger than the Lifshitz symmetry. While the Schr\" odinger
symmetry \cite{Hagen:1972pd, Niederer:1972zz, Mehen:1999nd, Nishida:2007pj} consists of time 
and space translations, 
spatial rotations, Galilean boosts,
dilatation or a scaling symmetry (mentioned above), a particle number symmetry and
in addition a special conformal transformation which appears only for $z=2$, the Lifshitz 
symmetry \cite{Lifshitz, Hornreich:1975zz} consists of only space-time translations, spatial 
rotations and a scaling symmetry.
Such non-relativistic symmetries arise in some strongly coupled condensed matter systems
near their quantum critical point \cite{Sachdev, Sachdev:2012dq}. In particular, the 
Schr\" odinger symmetry arises
in fermionic cold atom system at unitarity \cite{Eagles:1969zz, Leggett, Nozieres:1985zz} 
and the Lifshitz symmetry arises in some strongly
correlated electron system such as certain dimer model \cite{Rokhsar:1988zz} 
and also in some lattice 
models \cite{Fradkin, Vishwanath, Ardonne}.
The gravity models which realize these symmetries as isometries \cite{Balasubramanian:2008dm,
Son:2008ye, Kachru:2008yh} are of interest as
they provide the calculational tool to study these strongly coupled condensed matter
systems \cite{Hartnoll:2009sz, Herzog:2009xv, McGreevy:2009xe, Sachdev:2011wg} in 
the spirit of AdS/CFT correspondence \cite{Maldacena:1997re, Gubser:1998bc, Witten:1998qj}.    

The metric having Schr\" odinger symmetry as an isometry has the form,
\be\label{schro}
ds^2 = -\frac{2 dt^2}{u^{2z}} + \frac{-2d\xi dt + \sum_{i=1}^d (dx^i)^2 + du^2}{u^2},
\ee
whereas, the metric having Lifshitz symmetry as an isometry has the form,
\be\label{lif}
ds^2 = - \frac{dt^2}{u^{2z}} + \frac{\sum_{i=1}^d (dx^i)^2 + du^2}{u^2}
\ee
In \eqref{schro} $\xi$ is a space-like coordinate whose conjugate $i\partial/\partial \xi$ 
is an operator associated with the conserved particle number and so, $\xi$ is compact. 
$z$ is the dynamical critical exponent as mentioned above. $u$ in 
\eqref{schro} and \eqref{lif} is the radial coordinate, which in the boundary theory
is related to the energy parameter giving rise to the RG flow. It is clear from
\eqref{schro} that the metric has a scaling symmetry $t \to \lambda^{z} t$, $\xi \to \lambda^{2-z}\xi$,
$x^i \to \lambda x^i$ and $u \to \lambda u$. On the other hand the metric in \eqref{lif} has
a scaling symmetry $t \to \lambda^{z} t$, $x^i \to \lambda x^i$ and $u \to \lambda u$. Apart
from the above scaling symmetry the metric in \eqref{schro} also has space-time translation,
spatial rotation, particle number ($\xi$ translation) and boost ($x^i \to x^i - v^i t$,
$\xi \to \xi + v^i x^i - (1/2)v^2t$) symmetry. Furthermore, for $z=2$, the metric has in 
addition a special conformal symmetry. Thus the metric \eqref{schro} has Schr\" odinger 
symmetry as an isometry. Whereas the Lifshitz metric in \eqref{lif} has apart from the
above mentioned scaling symmetry, space-time translation and spatial rotation symmetry and thus
has Lifshitz symmetry as an isometry.

It is well-known in a `top-down' approach that Schr\" odinger metric \eqref{schro} can be 
obtained either by the so-called `Null Melvin Twist' \cite{Bergman:2001rw, Alishahiha:2003ru,
Gimon:2003xk, Hubeny:2005qu, Balasubramanian:2008dm} or by the TsT transformation \cite{Maldacena:2008wh}
on the standard
$p$-brane/M-brane solutions of string/M theory and then taking dimensional reduction.
However, the non-relativistic branes or the Schr\" odinger metric obtained this way are
not supersymmetric. Supersymmetric non-relativistic branes have been obtained in \cite{Donos:2009en,
Pal:2009np}. Lifshitz
metric \eqref{lif}, on the other hand, is not easy to obtain directly from string/M theory.
The gravity dual of Lifshitz metric was first obtained in \cite{Kachru:2008yh}. 
Lifshitz solution with a specific dynamical exponent, namely, $z=2$, has been obtained by AdS null
deformations of known string theory solution in \cite{Balasubramanian:2010uk}. It has been either 
generalized \cite{Donos:2010tu} or
embedded in a `bottom-up' approach in various gauged supergravities as well as in string/M
theory in \cite{Hartnoll:2009ns}. However, a direct `top-down' approach of obtaining Lifshitz 
metric was still missing. In 
\cite{Dey:2012tg, Dey:2012rs, Dey:2012fi}, it was shown how to obtain Lifshitz metric directly from 
certain intersecting brane solutions
of string/M theory by taking the near horizon limit and then reducing them to lower dimensions.  
It should be remarked here that the metrics obtained by these methods are not 
invariant under the Schr\" odinger or  Lifshitz symmetries, rather, are conformal to the Schr\" odinger
or Lifshitz metrics. The conformal factor is related to the so-called hyperscaling violation exponent
of the boundary theory \cite{Fisher:1986zz} and so the above mentioned metrics are called the
Schrodinger or Lifshitz metrics with hyperscaling violation. Such space-times and their properties have
been studied in \cite{Ogawa:2011bz, Huijse:2011ef, Dong:2012se, Kim:2012nb, Narayan:2012hk, Singh:2012un,
Gath:2012pg}.           

In this paper we argue that {\it the single brane or the intersecting brane solutions of
string/M theory which are known to lead to AdS geometry (directly or upto a conformal transformation)
in the near horizon limit also give 
Schr\" odinger/Lifshitz dual space-times (without or with hyperscaling violation) upon using appropriate 
solution generating technique and dimensional reduction.}   
In order to have this, the dimensions of the AdS space must be greater than 2. Also, 
both the Schr\" odinger
and Lifshitz space-times obtained this way are supersymmetric as they are obtained from BPS solutions
of string/M theory using solution generating transformation (without breaking supersymmetry). We
illustrate this with several examples. Among the single branes M2- and M5-branes of M-theory lead directly 
to AdS$_4$ and AdS$_7$ geometry respectively in the near horizon limit. Similarly, D3-brane of string theory
leads to AdS$_5$ geometry directly. Other D$(p+1)$-branes of string theory lead to AdS$_{p+3}$ geometry upto 
a conformal transformation except for $p=4$. We will see that in all these cases the solutions lead to
Schr\" odinder/Lifshitz dual space-times.
The intersecting branes of string/M theory which give AdS geometry have been given in \cite{Boonstra:1997dy,
Boonstra:1998yu}.
Among them the ones which yield AdS$_3$ (AdS$_2$ is excluded as we mentioned) geometry are M2 $\perp$ M5 and 
M5 $\perp$ M5 $\perp$ M5 \cite{Klebanov:1996mh} of M-theory and D1 $\perp$ D5, D2 $\perp$ D4, D3 $\perp$ D3 and F $\perp$ NS5
\cite{Dey:2012tg, Dey:2012rs} of type II string theories (the solutions involving triple intersections 
of string theory branes are
not known explicitly and even if they give AdS geometry, they will either give AdS$_3$ or AdS$_2$ similar
to the solutions involving double intersections and therefore we will not consider them here to avoid 
repetitions). We will consider some of these cases to show how they lead to Schr\" odinger/Lifshitz dual
space-times (except F $\perp$ NS5). Now starting from the AdS geometry we
first generate a pp-wave along one of the original brane directions for single brane or one of the original
common brane directions for the intersecting branes by standard method \cite{Russo:1996if}. We then express the resulting
solution in the light-cone coordinates and by further taking double Wick 
rotations involving the light cone  coordinates we generate the Schr\" odinger space-time (upon dimensional 
reductions). On the other hand to obtain Lifshitz
space-time, we either dimensionally reduce the solution (for M-theory branes) in Poincare coordinates 
along the wave direction
or take T-duality (for string theory branes) along the same direction (and making further dimensional
reductions). This way we generate 
Schr\" odinger/Lifshitz dual space-times starting from the same solution of string/M theory. The Schr\" odinger
or Lifshitz space-times obtained this way can have hyperscaling violations in some cases. We will also
show that the dynamical critical exponents of these two space-times always add upto 2. As the 
Schr\" odinger/Lifshitz space-times are obtained from BPS solutions of string/M theory, they preserve some
fraction of space-time supersymmetry. Hyperscaling violating Lifshitz solutions have also been obtained from 
Schr\" odinger space-time and branes with waves in \cite{Gath:2012pg}.

This paper is organized as follows. In section 2, we give the general argument of obtaining 
Schr\" odinger/Lifshitz dual space-times from the AdS geometry. The examples illustrating this are
given in the next two sections. In section 3, we discuss the cases of single branes i.e., M2-, M5-
and D$(p+1)$-branes. In section 4, we discuss the cases of intersecting branes including M2 $\perp$ M5,
M5 $\perp$ M5 $\perp$ M5 of M theory and D1 $\perp$ D5, D2 $\perp$ D4 and F $\perp$ NS5 of string
theory. Finally, we conclude in section 5.

\section{From AdS to Schr\" odinger/Lifshitz dual space-times}

It is well-known that the non-dilatonic branes of string/M theory lead to AdS geometry in the 
near horizon limit. So, for example, M2- and M5-branes of M theory give AdS$_4$ and AdS$_7$ respectively, 
whereas D3-brane of string theory gives AdS$_5$ space-times in the near horizon limit apart from some 
spherical part which we will not need in our discussion here. However, the other D$(p+1)$ branes ($p\neq 2$)
of string theory do not give AdS geometry in the near horizon limit directly, but they give 
geometries which are conformal to AdS$_{p+3}$ except for $p=4$. Other BPS branes like F-string or 
NS5-branes do not give AdS geometry and so, the only single branes that are relevant for our
purpose here are the M2-, M5- and D$(p+1)$-branes (for $p \neq 4$). Intersecting branes can also lead to
AdS geometries, the first example being M2-brane intersecting with M5-brane on a common string,
M2 $\perp$ M5, which gives AdS$_3$ geometry in the near horizon limit. This solution preserves 1/4
supersymmetry. Another example of intersecting
M-branes which gives AdS$_3$ geometry is triple intersections, namely, three M5-branes intersecting
on a string and pairwise intersecting on 3-branes, M5 $\perp$ M5 $\perp$ M5 \cite{Klebanov:1996mh}. 
This solution preserves
1/8 supersymmetry. There are other intersecting solutions of M-branes, for example, M2 $\perp$ M2 
$\perp$ M2, M2 $\perp$ M2 $\perp$ M5 $\perp$ M5, which presererve some supersymmetries and give AdS
geometries \cite{Boonstra:1998yu}, but these latter solutions give AdS$_2$ which does not contain 
any brane direction and
therefore are not suitable for generating Schr\" odinger/Lifshitz dual space-times. Intersecting
solutions of string theory can also lead to AdS geometry in the near horizon limit. Some of the
double intersecting solutions are discussed in \cite{Dey:2012tg, Dey:2012rs}. Among them those which give AdS$_3$ are
D1 $\perp$ D5, D2 $\perp$ D4, D3 $\perp$ D3 and F $\perp$ NS5. Note that although F-string and
NS5-brane individually does not give AdS geometry their intersection gives AdS$_3$. We will discuss
D1 $\perp$ D5, D2 $\perp$ D4 to show how they give rise to Schr\" odinger/Lifshitz
dual space-times. We also discuss the case F $\perp$ NS5 as an exception. We point out that even if
it gives AdS geometry, it does not yield Schr\" odinger/Lifshitz dual space-times.  We do not consider 
triple intersections in string theory for the reasons mentioned earlier.

In this section we will give the general arguments to show how starting from AdS solutions (obtained from
various single brane or double/triple intersecting solutions of string/M theory) we can generate
Schr\" odinger/Lifshitz dual space-times by using some solution generating technique. Suppose
we start from any such solution of string/M theory described in the first paragraph whose near
horizon limit gives AdS geometry either directly or upto a conformal transformation. Then the
near horizon metric can be written as,
\be\label{adsgeom}
ds^2 = u^{\beta}\left[\frac{-dt^2 + (dx^1)^2 + \sum_{i=2}^{d} (dx^i)^2 + du^2}{u^2}\right]
\ee
where we have put all the charges associated with the branes to unity for convenience.
In the above $u$ is a radial coordinate which is related to the original radial coordinate
$r$, transverse to the branes, by some coordinate transformation. The overall $u^{\beta}$, 
where $\beta$ is a constant, 
indicates that the metric is conformal to AdS$_{d+2}$ space. Note that we have isolated one of the
brane directions $x^1$ along which waves will be generated. Also, we have ignored the spherical 
part and some Euclidean part as they will not play any role in our discussion here. 

In order to obtain Schr\" odinger space-time from here we will generate waves along one of the
brane directions $x^1$ (say). The standard
way to generate waves along a brane direction is to first write the black solution and then
boost the solution along that direction and then take double scaling limit \cite{Russo:1996if}. The resultant 
solution is an extremal brane solution with waves along $x^1$. Applying this method, the above
metric \eqref{adsgeom} reduces to,
\be\label{adsgeomwave}
ds^2 = u^{\beta}\left[\frac{-dt^2 + (dx^1)^2 + (H-1)(dt-dx^1)^2 + \sum_{i=2}^{d} (dx^i)^2 + du^2}{u^2}\right]           
\ee
where $H = (1+u^{\alpha})$ is a harmonic function and $\alpha$ is another constant. Now defining
the light-cone coordinates by 
\be\label{newcoord} 
t_{\rm  new} = \frac{1}{\sqrt 2} (t+x^1), \quad \xi =   \frac{1}{\sqrt 2} (t-x^1)
\ee
we rewrite \eqref{adsgeomwave} as,
\be\label{adsgeomwavelc}
ds^2 =  u^{\beta}\left[\frac{-2 dt d\xi + 2 u^{\alpha} d\xi^2 + \sum_{i=2}^{d} (dx^i)^2 + du^2}{u^2}\right]
\ee
Note that in writing \eqref{adsgeomwavelc} we have replaced $t_{\rm new} \to t$. Now if we make a double
Wick rotation $t \to i\xi$ and $\xi \to -it$, the metric \eqref{adsgeomwavelc} takes the form
\be\label{wick}
ds^2 =   u^{\beta}\left[-\frac{2 dt^2}{u^{2-\alpha}} + \frac{-2 d\xi dt + \sum_{i=2}^{d} (dx^i)^2 + du^2}{u^2}\right]
\ee
Now comparing \eqref{wick} with the Schr\" odinger metric given in \eqref{schro} we find that apart from
overall $u^{\beta}$ factor the metric has exactly the same form with the dynamical critical exponent
$z = 1-\alpha/2$. The overall factor represents that the whole metric actually transforms under the scaling
and the parameter $\beta$ is related to the so-called hyperscaling violation exponent. This therefore shows
that how starting from AdS geometry (obtained from the near horizon limit of brane solutions of string/M theory)
we can get Schr\" odinger space-times with (for $\beta \neq 0$) or without (for $\beta=0$) hyperscaling
violation.

Now in order to get Lifshitz space-times we note that we can rewrite the metric in \eqref{adsgeomwave} as,
\be\label{prelif}
ds^2 = u^{\beta}\left[\frac{-H^{-1} dt^2 + H\left((1-H^{-1})dt - dx^1\right)^2 + \sum_{i=2}^{d} 
(dx^i)^2 + du^2}{u^2}\right] 
\ee
Now if the above metric is obtained from M-theory solution we can dimensionally reduce it along $x^1$
to go to a string theory solution which can be written as,
\be\label{prelif1}
ds^2 = u^{\beta+\gamma}\left[\frac{-H^{-1} dt^2 + \sum_{i=2}^{d} (dx^i)^2 + du^2}{u^2}\right] 
\ee 
where $\gamma$ is another constant. Note that the second term in \eqref{prelif} is absent in the
reduced solution and a gauge field $A_0$ will be generated in string theory solution. Thus
here the dimensionality of the solution or the boundary theory is reduced by one. The solution
\eqref{prelif1} in the near horizon limit takes the form,
\be\label{prelif2}
ds^2 = u^{\beta+\gamma}\left[- \frac{dt^2}{u^{\alpha+2}} + \frac{\sum_{i=2}^{d} (dx^i)^2 + du^2}{u^2}\right] 
\ee
In the above we have used the fact that in the near horizon limit $H = 1 + u^{\alpha} \approx u^\alpha$.
Actually, as will see in the specific examples in the next sections that the near horizon limit implies
$u \to \infty$ ($u \to 0$) and in that case $\alpha$ is positive (negative) and so, we always 
have $u^\alpha \gg 1$. 
On the other hand if the metric \eqref{prelif} is obtained from a string theory solution, we can take
T-duality along $x^1$ and the resulting solution in that case can be written as,
\be\label{prelif3}
ds^2 = u^{\beta}\left[- \frac{dt^2}{u^{\alpha+2}} + \frac{\sum_{i=1}^{d} (dx^i)^2 + du^2}{u^2}\right] 
\ee
In this case if we start from a type IIA solution we will end up with a type IIB solution and 
vice-versa. However, the dimensionality of the solution or the boundary theory does not change.
Thus comparing \eqref{prelif2} and \eqref{prelif3} with \eqref{lif}, we find that in either 
case we get Lifshitz solution without or with
hyperscaling violation with the dynamical critical exponent $z=1+\alpha/2$. Now since for the 
Schr\" odinger
solution we get $z=1-\alpha/2$ we thus find that the sum of the dynamical critical exponents of the
Schr\" odinger/Lifshitz dual space-times obtained this way add upto 2.

Before we close this section, we would like to emphasize that the reason we were able to obtain
Schr\" odinger/Lifshitz dual space-times from a given AdS solution of string/M theory lies in
our ability to write the AdS + pp-wave solution in two different ways as given in eqs. \eqref{adsgeomwave}
and \eqref{prelif} above. In the first case we obtain Schr\" odinger metric (upto a conformal factor)
by going to the light-cone coordinates and then making a double Wick rotation, whereas, in the second case
we obtain Lifshitz metric (upto a conformal factor) by either dimensionally reducing the solution
(for M-theory) or taking T-duality transformation (for string theory). It is well known that AdS solution
of string/M theory can be deformed more generally by some non-normalizable terms which include the matter
sector of the theory and in particular, certain AdS null deformations can lead to Lifshitz solutions 
\cite{Balasubramanian:2010uk}. 
Although these deformations are more general than the pp-wave deformations we consider in this paper,
they may not lead to Schr\" odinger/Lifshitz dual space-times in general. In other words,
for the occurrence of the dual space-times, we need to consider a special deformation of the AdS solution,
namely, the deformation by pp-wave. Also note that since $\alpha$ as given after eq.\eqref{adsgeomwave}
is a positive number, the dynamical critical exponent $z=1-\alpha/2$ for the Schr\" odinger case can 
become negative.
In fact $z$ can become negative even for some Lifshitz (with hyperscaling violation) solutions as given in 
section 4.2 below. This may seem surprising but in the boundary theory this simply means that the dynamical
system near the critical point has smaller relaxation time (instead of longer relaxation time for $z>0$)
or in this case there will be a critical speeding up (instead of slowing down) of the system.
This is known to occur for some random-cluster model \cite{Deng:2007dh} and also in 3-dimensional 
single cluster O(4) sigma model \cite{Rummukainen}. Note that negative $z$ implies that the time and the
space coordinates scale in the opposite way, however, this does not pose any problem to consider the
theory at finite temperature for Lifshitz case. In order to add a temperature we need to put a suitable 
Schwarzschild factor, as usual, in $dt^2$ and $dr^2$ terms since the Lifshitz metric is diagonal, but we do 
not consider this case here. The finite temperature solution in this case can be shown to be stable even
for $z<0$ as long as the hyperscaling parameter satisfies $\theta > d$. Indeed this happens for some cases
considered later in section 4. However, how to add temperature for the Schr\" odinger case is not obvious 
at all because of the presence of a cross-term $dt d\xi$ in the metric, but this problem is in no way related
to the negative value of the dynamical exponent. Since the particular Schr\" odinger space-time considered here
is obtained by generating a wave along one of the brane directions and that involves taking a zero temperature 
and infinite boost limit (such that their product remains constant) of the boosted black brane solutions,
it is not clear how to further add temperature to these solutions. 
       
Thus we have shown how starting from AdS geometry obtained from some string/M theory solutions we 
can generate both the Schr\" odinger and the Lifshitz space-times without and with hyperscaling violations.
Here we have argued in generality in a schematic fashion and the details will be given as we discuss 
various specific cases in the next two sections.

\section{Schr\" odinger/Lifshitz from single brane solutions}

In this section we will consider single brane solutions, namely, M2-, M5-branes of M-theory
and D$(p+1)$-branes of string theory which are known to lead to AdS geometry in the near
horizon limit either directly or
upto a conformal transformation. We will show how they lead to Schr\" odinger/Lifshitz dual
space-times upon using some solution generating techniques.

\subsection{M2-brane}

M2-brane solution has the form,
\bea\label{m2}
ds^2 &=& H_2^{-\frac{2}{3}}\left(-dt^2 + (dx^1)^2 + (dx^2)^2\right) + H_2^{\frac{1}{3}}
\left(dr^2 + r^2 d\Omega_7^2\right)\nn
A_{[3]} &=& H_2^{-1} dt \wedge dx^1 \wedge dx^2
\eea
Here $H_2$ is a harmonic function given as $H_2(r) = 1 + Q_2/r^6$, where $Q_2$ is the charge associated
with the M2-brane and $A_{[3]}$ is the 3-form gauge field which couples to M2-brane. In the near horizon
limit, $r \to 0$, the above metric reduces to
\be\label{nhm2} 
ds^2 = \frac{r^4}{Q_2^{\frac{2}{3}}}\left(-dt^2 + (dx^1)^2 + (dx^2)^2\right) + \frac{Q_2^{\frac{1}{3}}}{r^2}
\left(dr^2 + r^2 d\Omega_7^2\right)
\ee
Now taking $r \to 1/r$ and defining a new variable by the relation $u^2=r^4$, we can rewrite \eqref{nhm2}
and the gauge field in \eqref{m2} as,
\bea\label{nhm2new}
ds^2 &=& Q_2^{\frac{1}{3}}\left[\frac{-dt^2 + (dx^1)^2 + (dx^2)^2 + \frac{1}{4} du^2}{Q_2 u^2} + d\Omega_7^2\right]\nn
A_{[3]} &=& \frac{1}{Q_2 u^3} dt \wedge dx^1 \wedge dx^2
\eea
This is the standard AdS$_4$ $\times$ S$^7$ metric which comes from M2-brane in the near horizon limit.
Now as mentioned in the previous section, to obtain the Schr\" odinger metric, we generate a pp-wave along
$x^1$ direction by the standard technique. The resultant metric then takes the form,
\be\label{m2wave}
ds^2 = Q_2^{\frac{1}{3}}\left[\frac{-dt^2 + (dx^1)^2 +(H_1-1)(dt-dx^1)^2 + (dx^2)^2 + \frac{1}{4} du^2}{Q_2 u^2} 
+ d\Omega_7^2\right]
\ee
where $H_1 = 1 + Q_1/r^6$ is another harmonic function and $Q_1$ is the asymptotic momentum carried by the wave.
As before taking the coordinate transformation $r \to 1/r$ and using the
new variable $u$, the harmonic function takes the form $H_1 = 1 + Q_1 u^3$. Note here that going
to the near horizon means $u \to \infty$. Writing the metric in \eqref{m2wave} and the form-field
in \eqref{nhm2new} in the light-cone coordinates as defined in the previous section we have,
\bea\label{nhm2lc}
ds^2 &=& Q_2^{\frac{1}{3}}\left[\frac{-2dtd\xi + 2 Q_1 u^3 d\xi^2 + (dx^2)^2 + \frac{1}{4} Q_2du^2}{Q_2 u^2} + d\Omega_7^2\right]\nn
A_{[3]} &=& -\frac{1}{Q_2 u^3} dt \wedge d\xi \wedge dx^2
\eea 
Now dimensionally reducing the solution on S$^7$ and taking the double Wick rotation $t \to i\xi$, $\xi \to -it$,
the above solution takes the form,
\bea\label{nhm2lcsch}
ds^2 &=& Q_2^{\frac{3}{2}}\left[-2\frac{Q_1}{Q_2}u dt^2 + \frac{-2d\xi dt + (dx^2)^2 + \frac{1}{4} Q_2du^2}{Q_2 u^2} \right]\nn
A_{[3]} &=& \frac{1}{Q_2 u^3} dt \wedge d\xi \wedge dx^2
\eea
Comparing the metric in \eqref{nhm2lcsch} with \eqref{schro}, we find that the metric has a Schr\" odinger symmetry
with dynamical critical exponent $z=-1/2$, spatial dimension of the boundary theory $d=1$ and no hyperscaling violation. 

Now to obtain Lifshitz metric, we rewrite the metric in \eqref{m2wave}, as mentioned in section 2, as
\be\label{m2lif1}
 ds^2 = Q_2^{\frac{1}{3}}\left[\frac{- H_1^{-1} dt^2 + H_1\left((1-H_1^{-1})dt - dx^1\right)^2 + (dx^2)^2 + \frac{1}{4}Q_2 du^2}{Q_2 u^2} 
+ d\Omega_7^2\right]
\ee
Reducing the above solution along the wave direction, i.e. along $x^1$, we obtain a string theory solution
which has the form,
\bea\label{stringliffd0}
ds^2 &=& Q_1^{\half} u^{\half} \left[-\frac{dt^2}{Q_1 Q_2 u^5} + \frac{(dx^2)^2 + \frac{1}{4}Q_2 du^2}{Q_2 u^2} + d\Omega_7^2\right]\nn
e^{2\phi} &=& \frac{Q_1^{\frac{3}{2}}}{Q_2} u^{\frac{3}{2}}\nn
B_{[2]} &=& \frac{1}{Q_2u^3} dt \wedge dx^2, \qquad A_{[1]}\,\,=\,\, \frac{1}{Q_1 u^3} dt
\eea
where we have written the metric in the string frame and used $H_1 \approx Q_1 u^3$. It can be easily seen that indeed
the above metric has Lifshitz symmetry with hyperscaling violation. 
Actually we recognize the solution \eqref{stringliffd0} to be the
near horizon limit of F-D0 solution discussed in \cite{Dey:2012tg}. This solution has, as discussed
in \cite{Dey:2012tg}, dynamical critical exponent $z=5/2$, the spatial dimension of boundary theory
$d=1$ and the hyperscaling violation exponent $\theta=1/2$.

We thus obtain both Schr\" odinger and Lifshitz space-times starting from the AdS$_4$ solution
which is the near horizon limit of M2-brane solution. The dynamical critical exponents of these
two space-times are $z=-1/2$ and $z=5/2$, which add upto 2. Note that since M2-brane is 
a BPS solution
both the Schr\" odinger and the Lifshitz solution obtained this way are supersymmetric.

\subsection{M5-brane}

Here we proceed exactly as in M2-brane case discussed in the previous subsection. The M5-brane
solution has the form,
\bea\label{m5}
ds^2 &=& H_2^{-\frac{1}{3}}\left(-dt^2 + (dx^1)^2 + \sum_{i=2}^5 (dx^i)^2\right) + H_2^{\frac{2}{3}}
\left(dr^2 + r^2 d\Omega_4^2\right)\nn
A_{[6]} &=& H_2^{-1} dt \wedge dx^1 \ldots \wedge dx^5
\eea 
In this case the harmonic function $H_2$ is given as $H_2(r) = 1 + Q_2/r^3$, where $Q_2$
is the charge associated with M5-brane. Also M5-brane couples to a 6-form gauge field 
given in \eqref{m5}. In the near horizon limit $r \to 0$, M5-brane solution takes the form,
\bea\label{m5nh}
ds^2 &=& \frac{r}{Q_2^{\frac{1}{3}}}\left(-dt^2 + (dx^1)^2 + \sum_{i=2}^5 (dx^i)^2\right) + 
\frac{Q_2^{\frac{2}{3}}}{r^2}\left(dr^2 + r^2 d\Omega_4^2\right)\nn
A_{[6]} &=& \frac{r^3}{Q_2} dt \wedge dx^1 \ldots \wedge dx^5
\eea 
Now taking the coordinate transformation $ r \to 1/r$ and defining a new coordinate by
$u^2 =r$, we can rewrite \eqref{m5nh} as,
\bea\label{m5nhnew}
ds^2 &=& Q_2^{\frac{2}{3}}\left[\frac{-dt^2 + (dx^1)^2 + \sum_{i=2}^5 (dx^i)^2 + 4 Q_2 du^2}{Q_2u^2} 
+ d\Omega_4^2\right]\nn
A_{[6]} &=& \frac{1}{Q_2 u^6} dt \wedge dx^1 \ldots \wedge dx^5
\eea 
The metric in \eqref{m5nhnew} has the standard AdS$_7$ $\times$ S$^4$ structure obtained from
the near horizon limit of M5-brane. Now in order to obtain Schr\" odinger solution we generate
pp-waves along $x^1$ direction by the standard procedure and then the metric will be given by,
\be\label{m5wave}
ds^2 = Q_2^{\frac{2}{3}}\left[\frac{-dt^2 + (dx^1)^2 +(H_1-1)(dt-dx^1)^2 + \sum_{i=2}^5(dx^i)^2 + 4Q_2 du^2}{Q_2 u^2} 
+ d\Omega_4^2\right]
\ee
where $H_1 = 1 +Q_1/r^3$ is another harmonic function, with $Q_1$, the asymptotic momentum carried by the wave.
Writing in terms of variable $u$, the harmonic function has the form $H_1 = 1 + Q_1 u^6$. As before here also
going to the near horizon means $u \to \infty$. In light cone coordinates the metric \eqref{m5wave} and the
form field in \eqref{m5nhnew} take the forms,
\bea\label{nhm5lc}
ds^2 &=& Q_2^{\frac{2}{3}}\left[\frac{-2dtd\xi + 2 Q_1 u^6 d\xi^2 + \sum_{i=2}^5(dx^i)^2 + 4 Q_2du^2}{Q_2 u^2} + d\Omega_4^2\right]\nn
A_{[6]} &=& -\frac{1}{Q_2 u^6} dt \wedge d\xi \wedge dx^2  \ldots \wedge dx^5
\eea 
Dimensionally reducing the solution on S$^4$ and taking the double Wick rotation as before
$t \to i\xi$ and $\xi \to -it$, we get,
\bea\label{nhm5lcwr}
ds^2 &=& Q_2^{\frac{6}{5}}\left[-2\frac{Q_1}{Q_2} u^4 dt^2 + \frac{-2d\xi dt + \sum_{i=2}^5 (dx^i)^2 + 4 Q_2 du^2}{Q_2 u^2}\right]\nn
A_{[6]} &=& \frac{1}{Q_2 u^6} dt \wedge d\xi \wedge dx^2 \ldots \wedge dx^5
\eea
Comparing the above metric with the Schr\" odinger metric given in \eqref{schro}, we find that
this metric has a Schr\" odinger symmetry with $z=-2$, $d=4$ and no hyperscaling violation.  

Now again in order to obtain Lifshitz metric we rewrite \eqref{m5wave} as,
\be\label{m5lif1}
 ds^2 = Q_2^{\frac{2}{3}}\left[\frac{- H_1^{-1} dt^2 + H_1\left((1-H_1^{-1})dt - dx^1\right)^2 + 
\sum_{i=2}^5(dx^i)^2 + 4 Q_2 du^2}{Q_2 u^2} 
+ d\Omega_4^2\right]
\ee 
Dimensionally reducing the solution along the wave direction, i.e., along $x^1$, we obtain
the following string theory solution,
\bea\label{stringlifd0d4}
ds^2 &=& Q_1^{\half}Q_2^{\half} u^2 \left[-\frac{dt^2}{Q_1 Q_2 u^8} + \frac{\sum_{i=2}^5(dx^i)^2 + 4Q_2 du^2}{Q_2 u^2} + d\Omega_4^2\right]\nn
e^{2\phi} &=& \frac{Q_1^{\frac{3}{2}}}{Q_2^{\half}} u^6\nn
A_{[1]} &=& \frac{1}{Q_1u^6} dt , \qquad A_{[5]}\,\,=\,\, \frac{1}{Q_2 u^6} dt \wedge dx^2 \ldots \wedge dx^5
\eea
where we have used $H_1 \approx Q_1u^6$ and the metric is written in the string frame. By comparing with
the Lifshitz metric \eqref{lif}, we immediately recognize that the metric in \eqref{stringlifd0d4} has a Lifshitz 
symmetry with hyperscaling violation. In fact we notice that the above solution is nothing but the near horizon 
limit of D0-D4 solution discussed in \cite{Dey:2012rs}. As found there D0-D4 solution in the near horizon limit indeed has
hyperscaling violating Lifshitz symmetry with dynamical critical exponent $z=4$, spatial dimension of the
boundary theory $d=4$ and the hyperscaling violation exponent $\theta=2$.

Thus we again found Schr\" odinger/Lifshitz space-times from AdS$_7$ solution which is the near horizon geometry
of M5-brane. Here we note that the dynamical critical exponents of Schr\" odinger/Lifshitz space-times are
given as $z = -2$ and $z=4$ respectively and so they again add upto 2 as expected.

\subsection{D(p+1)-branes}
   
The D$(p+1)$-brane solution of type II string theory has the form,
\bea\label{dp1brane}
ds^2 &=& H_2^{-\half}\left(-dt^2 + (dx^1)^2 + \sum_{i=2}^{p+1}(dx^i)^2\right) + H_2^{\half}\left(dr^2 + r^2 d\Omega_{7-p}^2\right)\nn
e^{2\phi} &=& H_2^{\frac{2-p}{2}}\nn
A_{[p+2]} &=& H_2^{-1} dt \wedge dx^1 \ldots \wedge dx^{p+1}
\eea
Here the metric is written in the string frame. The harmonic function $H_2$ has the form $H_2 = 1 + Q_2/r^{6-p}$. Note
that we have isolated one of the brane directions ($x^1$) along which pp-waves will be generated. This is the reason
$p \neq 0$ and in fact we have $ 1 \leq p \leq 5$. $Q_2$ is the charge associated with D$(p+1)$-brane. $\phi$ is the 
dilaton and $A_{[p+2]}$ is a $(p+2)$-form field which couples to D$(p+1)$-brane. In the near horizon limit the above metric
takes the form,
\be\label{dp1nh}
ds^2 = Q_2^{\half} r^{\frac{p-2}{2}}\left[\frac{r^{4-p}}{Q_2}\left(-dt^2 + (dx^1)^2 + \sum_{i=2}^{p+1}(dx^i)^2\right) + \frac{dr^2}{r^2}
+ d\Omega_{7-p}^2\right]
\ee
Now going to a coordinate $r \to 1/r$ and introducing a new variable by the relation $u^2=r^{4-p}$ (for $p \neq 4$), we can
rewrite the metric \eqref{dp1nh} along with the other fields given in \eqref{dp1brane} as,
\bea\label{dp1nhfull}
ds^2 &=& Q_2^{\half} u^{\frac{2-p}{4-p}}\left[\frac{-dt^2 + (dx^1)^2 + \sum_{i=2}^{p+1}(dx^i)^2 + 
\frac{4}{(4-p)^2}Q_2 du^2}{Q_2 u^2} + d\Omega_{7-p}^2\right]\nn
e^{2\phi} &=& Q_2^{\frac{2-p}{2}} u^{\frac{(2-p)(6-p)}{(4-p)}}\nn
A_{[p+2]} &=& \frac{1}{Q_2 u^{\frac{2(6-p)}{(4-p)}}} dt \wedge dx^1 \ldots \wedge dx^{p+1}
\eea
Here the metric has AdS$_{p+3}$ $\times$ S$^{7-p}$ structure for $1 \leq p \leq 5$ except for
$p=4$ (or D5-brane) upto a conformal factor ($u^{(2-p)/(4-p)}$). Note that the conformal factor
actually vanishes for $p=2$ or D3-brane as is well-known. For $p=4$ or D5-brane, we do not
get AdS in the near horizon limit, rather we get ${\cal M}_7$ $\times$ S$^3$ upto a conformal
factor, where ${\cal M}_7$ represents the seven dimensional Minkowski space. Now we will show
how starting from this AdS$_{p+3}$ geometry we can generate Schr\" odinger/Lifshitz space-times
by some solution generating transformation as was done for M2- and M5-brane cases. 

In order to get Schr\" odinger space-times we generate pp-waves, by standard technique \cite{Russo:1996if}, along 
$x^1$ direction as before and thus we obtain the metric
\be\label{dp1wave}
ds^2 = Q_2^{\half} u^{\frac{2-p}{4-p}}\left[\frac{-dt^2 + (dx^1)^2 + (H_1-1) (dt-dx^1)^2 + \sum_{i=2}^{p+1}(dx^i)^2 + 
\frac{4}{(4-p)^2}Q_2 du^2}{Q_2 u^2} + d\Omega_{7-p}^2\right]  
\ee
where $H_1 = 1 + Q_1/r^{6-p}$ is a harmonic function with $Q_1$, the asymptotic momentum carried by the wave.
By first changing $r \to 1/r$ and then defining $u^2 = r^{4-p}$, the harmonic function can be witten as
$H_1 = 1 + Q_1 u^{2(6-p)/(4-p)}$. Note that for $p < 4$, the near horizon limit $r \to \infty$ implies 
$u \to \infty$, but
for $p > 4$, $r \to \infty$ implies $u \to 0$. However in both cases, in the near horizon limit
$H_1 \approx Q_1 u^{2(6-p)/(4-p)}$. Now in the light-cone coordinates defined earlier, the above metric
\eqref{dp1wave} can be written as,
 \be\label{dp1wavelc}
ds^2 = Q_2^{\half} u^{\frac{2-p}{4-p}}\left[\frac{-2dtd\xi + 2Q_1 u^{\frac{2(6-p)}{(4-p)}} d\xi^2 + \sum_{i=2}^{p+1}(dx^i)^2 + 
\frac{4}{(4-p)^2}Q_2 du^2}{Q_2 u^2} + d\Omega_{7-p}^2\right]  
\ee
Dimensionally reducing the solution on S$^{7-p}$ and expressing the resultant
metric in the Einstein frame and then further taking the double Wick rotation ($t \to i\xi$ and
$\xi \to -it$), the metric \eqref{dp1wavelc} and the other fields in \eqref{dp1nhfull} 
take the forms,
\bea\label{dp1sch}
ds^2 &=& Q_2^{\frac{p+2}{p+1}} u^{\frac{2(p-2)^2}{(p-4)(p+1)}}\left[-2\frac{Q_1}{Q_2} u^{\frac{4}{4-p}} dt^2 + \frac{-2d\xi dt
+ \sum_{i=2}^{p+1}(dx^i)^2 + \frac{4}{(4-p)^2} Q_2 du^2}{Q_2 u^2}\right]\nn
e^{2\phi} &=&  Q_2^{\frac{2-p}{2}} u^{\frac{(2-p)(6-p)}{(4-p)}}\nn
A_{[p+2]} &=& \frac{1}{Q_2 u^{\frac{2(6-p)}{(4-p)}}} dt \wedge d\xi \wedge dx^2 \ldots \wedge dx^{p+1}
\eea 
Comparing the metric in \eqref{dp1sch} with the Schr\" odinger metric given in \eqref{schro}
we immediately notice that this metric has a Schr\" odinger symmetry with hyperscaling violation.
Under the scaling symmetry $t \to \lambda^{-\frac{2}{4-p}}$, $\xi \to \lambda^{2+\frac{2}{4-p}} \xi$, 
$x^i \to \lambda x^i$ (for $i=2,\ldots,(p+1)$), $u \to \lambda u$, the metric in the square bracket
remains invariant. However, as there is a hyperscaling violation the full metric is not invariant
under the scaling, but changes as $ds \to \lambda^{(p-2)^2/((p-4)(p+1))} ds \equiv \lambda^{\theta/D} ds$,
where $D$ is kept arbitrary as in \cite{Kim:2012nb}.
We thus find that the metric in \eqref{dp1sch} has a hyperscaling violating Schr\" odinger symmetry
with the dynamical critical exponent $z= -2/(4-p)$, hyperscaling violation exponent $\theta/D = 
(p-2)^2/((p-4)(p+1))$ and the spatial dimension of the boundary theory $d = p$. We note that for $p=2$,
i.e. for D3-brane $\phi$ = constant and $\theta/D=0$ and for $p<4$, $\theta/D < 0$. Therefore, 
there is no hyperscaling violation for $p=2$ and hyperscaling violating exponent is negative 
for $p<4$. Similar observations were made \cite{Kim:2012nb} for the non-relativistic D$p$-branes obtained 
by the Null Melvin Twist \cite{Mazzucato:2008tr}.  

Now in order to get Lifshitz space-time we rewrite the metric with waves in \eqref{dp1wave}
as follows,
\be\label{dp1wavelif}
ds^2 = Q_2^{\half} u^{\frac{2-p}{4-p}}\left[\frac{-H_1^{-1} dt^2 + H_1\left((1-H_1^{-1})dt - dx^1\right)^2  + \sum_{i=2}^{p+1}(dx^i)^2 + 
\frac{4}{(4-p)^2}Q_2 du^2}{Q_2 u^2} + d\Omega_{7-p}^2\right]  
\ee
Since here we are dealing with string theory solutions, we will take T-duality along the wave direction, i.e., $x^1$
to obtain Lifshitz metric. After T-duality the only metric components that will change are $g_{tt}$, $g_{tx^1}$ and $g_{x^1x^1}$.
Denoting the new metric components by a `tilde', we have
\bea\label{metric} 
{\tilde g}_{tt} &=& g_{tt} - \frac{g_{tx^1}^2}{g_{x^1x^1}} \,\, = \,\, -\frac{Q_2^{\half} u^{\frac{2-p}{4-p}}}{Q_2 u^2} H_1^{-1}
\,\,=\,\, -\frac{Q_2^{\half} u^{\frac{2-p}{4-p}}}{Q_1Q_2 u^{\frac{4(5-p)}{4-p}}}\nn
{\tilde g}_{tx^1} &=& \frac{B_{tx^1}}{g_{x^1x^1}} \,\,=\,\, 0\nn
{\tilde g}_{x^1x^1} &=& \frac{1}{g_{x^1x^1}} \,\,=\,\, \frac{Q_2^{\half} u^{\frac{2-p}{4-p}}}{Q_1 u^2}
\eea
where we have used the metric \eqref{dp1wavelif}. We have also used $H_1 \approx Q_1 u^{2(6-p)/(4-p)}$ in the near
horizon limit. ${\tilde g}_{tx^1}$ vanishes because D$(p+1)$-brane solution with wave does not have an NSNS
$B$-field. However, in the T-dual solution a $B$-field will be generated because of a non-vanishing $g_{tx^1}$
component in \eqref{dp1wavelif}. Using \eqref{metric} and the other field configuration given in \eqref{dp1nhfull}
the complete T-dual solution can be written as,
\bea\label{dp1wavetdual}
ds^2 &=& Q_2^{\half} u^{\frac{2-p}{4-p}} \left[-\frac{dt^2}{Q_1Q_2 u^{\frac{4(5-p)}{4-p}}} + \frac{\frac{Q_2}{Q_1} (dx^1)^2 +
\sum_{i=2}^{p+1}(dx^i)^2 + \frac{4}{(4-p)^2}Q_2 \frac{du^2}{u^2}}{Q_2 u^2} + d\Omega_{7-p}^2\right]\nn
e^{2\phi} &=& \frac{Q_2^{\frac{3-p}{2}}}{Q_1} u^{\frac{(6-p)(1-p)}{(4-p)}}\nn
B_{[2]} &=& \frac{1}{Q_1 u^{\frac{2(6-p)}{4-p}}} dt \wedge dx^1,\qquad A_{[p+1]} \,\,=\,\,   
\frac{1}{Q_2 u^{\frac{2(6-p)}{4-p}}} dt \wedge dx^2 \wedge \ldots \wedge dx^{p+1}
\eea 
By comparing with the metric \eqref{lif}, we find that the metric in \eqref{dp1wavetdual} indeed has a Lifshitz 
symmetry with hyperscaling violation. In fact we recognize this solution as the  near horizon limit
of the 1/4 BPS, F-D$p$ solution found in \cite{Dey:2012tg}. Here F-string lies along $x^1$ and D$p$-brane lies
along $x^2,\,\ldots,\,x^{p+1}$. Note that the gauge fields differ by a sign from those in \cite{Dey:2012tg} due
to a slightly different convention we use here. As discussed in \cite{Dey:2012tg}, the metric in \eqref{dp1wavetdual} 
has a hyperscaling
violating Lifshitz symmetry with the dynamical critical exponent $z= 2(5-p)/(4-p)$, hyperscaling violation
exponent $\theta = p - (p-2)/(4-p)$ and the spatial dimension of the boundary theory $d = p+1$. Interestingly,
as noted also in \cite{Dey:2012tg}, for $p=2$, or D3-brane we have $\theta=p=d-1$. The entanglement
entropy of the boundary theory in that case is well-known to show a logarithmic violation of the area law and
therefore represents compressible metallic state with hidden fermi surface \cite{Ogawa:2011bz, Huijse:2011ef}.
The entanglement entropy for the F-D$p$ system has been calculated in \cite{Dey:2012hf}.
   
Thus we have shown that starting from the AdS$_{p+3}$ geometry (upto a conformal transformation) obtained
from the near horizon limit of D$(p+1)$-brane solutions of type II string theory, we can generate both the
Schr\" odinger as well as Lifshitz space-times by making use of some solution generating techniques. Note 
that since the critical dynamical exponent for the Schr\" odinger symmetry has the value $z= -2/(4-p)$
and that of the Lifshitz symmetry has the value $z=2(5-p)/(4-p)$, they indeed add upto 2 as we argued
in section 2.

Before we conclude this section, we would like to remark that, for the case of D$p$-branes, Schr\" odinger/Lifshitz 
space-times have also been obtained in \cite{Singh:2012un}. However, their
connections with AdS geometry was not clear there. Lifshitz geometry was obtained in \cite{Singh:2012un} by first 
taking a double scaling limit of the boosted black D$p$-brane (this is a standard procedure for 
generating pp-waves \cite{Russo:1996if}), then going to the light-cone coordinates and finally compactifying the solution along
the space-like light-cone direction. Note that in this procedure the solution becomes nine-dimensional and also the 
dimensionality of the boundary theory gets reduced by one. Since the light-cone coordinates involve one of the
brane directions, it is not clear why one should compactify that direction. However, we think that the proper way to
identify the Lifshitz geometry from D$p$-branes, as is done in this paper, is to remain in Poincare
coordinates and take a T-duality. This way there is no need to compactify one of the brane directions and the
dimensionality of the boundary theory does not get reduced. Also our method clarifies the connection of
Lifshitz space-times with AdS geometry as a deformation of the latter by the pp-waves and
a performance by T-duality. Note, however, that for M-theory solutions we performed a dimensional reduction
along the wave direction in order to obtain Lifshitz symmetry (there is no T-duality here). But since 
this extra dimension of M-theory
is related to the string coupling constant, this means that Lifshitz symmetry is manifest only at small
string coupling. When the coupling is large we have to lift the solution to M-theory and in that case we 
get an asymmetric Lifshitz scaling of this extra dimension. The Schr\" odinger space-time, on the other hand, 
was obtained in \cite{Singh:2012un},
by first going to the corresponding bubble solutions with a double Wick rotation, then
taking a double scaling limit on the boosted bubble solutions and finally going to the light-cone coordinates.
However, in this paper, we started from AdS solution and then deformed it by a pp-wave. We further introduced
the light-cone coordinates and then took the double Wick rotation. This way we obtained Schr\" odinger space-times
with hyperscaling violation. We also identified the hyperscaling violation exponent in this case, a feature
never mentioned in \cite{Singh:2012un}. We have seen that the dynamical critical exponents of the Schr\" odinger/Lifshitz dual
space-times add upto 2. This was also observed in \cite{Singh:2012un} in various cases of D$p$-branes. We have given a general
argument in section 2 explaining why this is so. Finally, in \cite{Singh:2012un}, it was shown how Lifshitz and Schr\" odinger
space-times arise from D$p$-brane solutions of string theory. Their origin was never really understood. In this
paper we have shown that whenever we get an AdS solution (directly or upto a conformal transformation) in string 
or M-theory, not necessarily only D$p$-branes, we can deform it (and apply other solution generating techniques)
to generate Schr\" odinger/Lifshitz dual space-times.    

\section{Schr\" odinger/Lifshitz from intersecting solutions}       

In this section we will show how one can get Schr\" odinger/Lifshitz dual space-times from some intersecting
solutions of M/string theory. The intersecting solutions we will consider give AdS geometry in the near horizon
limit. For M-theory it is known that there are two intersecting solutions which lead to AdS$_3$ geometry and
they are M2 $\perp$ M5 and M5 $\perp$ M5 $\perp$ M5 \cite{Boonstra:1998yu}. There are others which lead to 
AdS$_2$ geometry, but we will not
consider them for the reason mentioned earlier. For string theory solutions there are double and triple intersections
which give AdS geometry, but we will consider only double intersections since the explicit triple intersecting 
solutions of string theory are not known and they will mostly give lower dimensional AdS space like the double 
intersecting solutions, therefore, Schr\" odinger/Lifshitz space-times will be quite similar as the double intersecting 
solutions. So, for string theory, we will consider D1 $\perp$ D5 (type IIB), D2 $\perp$ D4 (type IIA)
and F $\perp$ NS5 (type IIA or IIB) solution all of which are known to lead to AdS$_3$ geometry in the near horizon 
limit. We will discuss the string theory solutions in brief as they give AdS$_3$ and the Schr\" odinger/Lifshitz
space-time they lead to are very similar to intersecting M-brane solutions. The low dimensional AdS space, namely,
AdS$_3$ along with waves and the resulting Schr\" odinger space have also been considered in \cite{Israel:2004vv}.  

\subsection{M2 $\perp$ M5 solution}

M2-brane intersecting with M5-brane on a string has a solution of the form,
\bea\label{m2m5}
ds^2 &=& H_2^{\frac{2}{3}}H_3^{\frac{1}{3}}\left[H_2^{-1}H_1^{-1}\left(-dt^2+(dx^1)^2\right) + H_2^{-1}\sum_{i=2}^5(dx^i)^2 
+H_3^{-1}(dx^6)^2 + dr^2 + r^2 d\Omega_3^2\right]\nn
A_{[3]} &=& H_3^{-1} dt \wedge dx^1 \wedge dx^6,\qquad A_{[6]}\,\,=\,\, H_2^{-1} dt \wedge dx^1 \wedge \ldots\wedge dx^5
\eea
Here the harmonic functions are given as $H_{2,3} = 1 + Q_{2,3}/r^2$ and $Q_{2,3}$ are the charges associated with
M5-brane and M2-brane respectively. Note that M2-brane lies along $x^1,\,x^6$ and M5-brane lies along $x^1,\,x^2,\,
\ldots, x^5$. In the near horizon limit $H_{2,3} \approx Q_{2,3}/r^2$ and the metric in \eqref{m2m5} takes the form,
\be\label{m2m5nh}
ds^2 = Q_2^{\frac{2}{3}}Q_3^{\frac{1}{3}}\left[\frac{r^2}{Q_2Q_3}\left(-dt^2+(dx^1)^2\right) + \frac{\sum_{i=2}^5(dx^i)^2}{Q_2} 
+\frac{(dx^6)^2}{Q_3} + \frac{dr^2}{r^2} + d\Omega_3^2\right]
\ee
Now defining a new coordinate by $u = 1/r$, we can write the full solution \eqref{m2m5} in the near horizon limit as,
\bea\label{m2m5nhu}
ds^2 &=& Q_2^{\frac{2}{3}}Q_3^{\frac{1}{3}}\left[\frac{-dt^2+(dx^1)^2 +Q_2Q_3du^2}{Q_2Q_3u^2} + \frac{\sum_{i=2}^5(dx^i)^2}{Q_2} 
+\frac{(dx^6)^2}{Q_3} + d\Omega_3^2\right]\nn
A_{[3]} &=& \frac{1}{Q_3 u^2}dt \wedge dx^1 \wedge dx^6,\qquad A_{[6]}\,\,=\,\, \frac{1}{Q_2u^2} dt \wedge dx^1 
\wedge \ldots\wedge dx^5
\eea
It is clear that the above metric has AdS$_3$ $\times$ E$^5$ $\times$ S$^3$ structure.

Now in order to get Schr\" odinger space-time we generate pp-waves along the common brane dicetion $x^1$ by 
the standard technique. The metric in \eqref{m2m5nhu} then takes the form,
\be\label{m2m5sch}
ds^2 = Q_2^{\frac{2}{3}}Q_3^{\frac{1}{3}}\left[\frac{-dt^2+(dx^1)^2 +(H_1-1)(dt-dx^1)^2+Q_2Q_3du^2}{Q_2Q_3u^2} + 
\frac{\sum_{i=2}^5(dx^i)^2}{Q_2} +\frac{(dx^6)^2}{Q_3} + d\Omega_3^2\right]
\ee
where $H_1 = 1 + Q_1/r^2$ is another harmonic function and $Q_1$ is the asymptotic momentum of the wave. Going
to the light cone coordinates and further taking double Wick rotation as before we arrive at the solution,
\bea\label{m2m5schfull}
ds^2 &=& Q_2^{\frac{2}{3}}Q_3^{\frac{1}{3}}\left[-\frac{2Q_1}{Q_2Q_3}dt^2 + \frac{-2d\xi dt +Q_2Q_3du^2}{Q_2Q_3u^2} + 
\frac{\sum_{i=2}^5(dx^i)^2}{Q_2} +\frac{(dx^6)^2}{Q_3} + d\Omega_3^2\right]\nn
A_{[3]} &=& \frac{1}{Q_3 u^2}dt \wedge d\xi \wedge dx^6,\qquad A_{[6]}\,\,=\,\, \frac{1}{Q_2u^2} dt \wedge d\xi 
\wedge dx^2 \wedge \ldots\wedge dx^5
\eea
The metric in \eqref{m2m5schfull} has a Schr\" odinger symmetry under the scaling $t \to \lambda^0 t$, $\xi \to
\lambda^2 \xi$ and $u \to \lambda u$. Note that the coordinates $x^2,\,\ldots,\,x^6$ do not scale and we can
compactify the solution on S$^3$ as well as E$^5$ to get a zero dimensional Schr\" odinger metric. This 
Schr\" odinger space-time has $d=0$, the dynamical critical exponent $z=0$ and no hyperscaling violation, 
i.e., $\theta=0$. 

Lifshitz space-time can be obtained by rewriting the metric \eqref{m2m5sch} as follows,
\be\label{m2m5lif}
ds^2 = Q_2^{\frac{2}{3}}Q_3^{\frac{1}{3}}\left[\frac{-H_1^{-1}dt^2+H_1\left((1-H_1^{-1})dt-dx^1\right)^2+Q_2Q_3du^2}{Q_2Q_3u^2} + 
\frac{\sum_{i=2}^5(dx^i)^2}{Q_2} +\frac{(dx^6)^2}{Q_3} + d\Omega_3^2\right]
\ee
Now compactifying along $x^1$, the wave direction, we get the ten dimensional string theory solution as,
\bea\label{m2m5liffull}
ds^2 &=& Q_1^{\half}Q_2^{\half}\left[-\frac{dt^2}{Q_1Q_2Q_3 u^4} + \frac{du^2}{u^2} + 
\frac{\sum_{i=2}^5(dx^i)^2}{Q_2} +\frac{(dx^6)^2}{Q_3} + d\Omega_3^2\right]\nn
e^{2\phi} &=& \frac{Q_1^{\frac{3}{2}}}{Q_2^{\half} Q_3}\nn
B_{[2]} &=& \frac{1}{Q_3 u^2}dt \wedge dx^6,\qquad A_{[5]}\,\,=\,\, \frac{1}{Q_2u^2} dt 
\wedge dx^2 \wedge \ldots\wedge dx^5, \qquad A_{[1]}\,\,=\,\,\frac{1}{Q_1 u^2} dt
\eea
Note that the metric has a scaling symmetry $t \to \lambda^2 t$, $u \to \lambda u$
and since the other spatial coordinates do not scale one can compactify the solution on 
S$^3$ $\times$ E$^5$ to get a zero dimensional Lifshitz metric. The dilaton is constant. 
The dynamical critical exponent
here is $z=2$ and hyperscaling violation exponent $\theta=0$. Now since the corresponding 
Schr\" odinger metric has dynamical critical exponent $z=0$ they indeed add upto 2. However,
note that since the other spatial dimensions do not scale, the zero dimensional Lifshitz
metric (consisting of the first two terms of the metric in \eqref{m2m5liffull}) can be
cast into an AdS$_2$ form by defining a new variable $\tilde u = u^2$. In this sense the zero
dimensional Lifshitz is kind of trivial. 

\subsection{More Lifshitz-like space-times}

As the solution \eqref{m2m5liffull} is obtained from the dimensional reduction (along the wave direction)
of the near horizon limit of `M2 $\perp$ M5 + wave' solution of M-theory, it must correspond to
the near horizon limit of F $\perp$ D4 $\perp$ D0 solution of string theory. Indeed one can check that
this is the case by explicitly constructing this solution (from the dimensional reduction of the complete
`M2 $\perp$ M5 + wave' solution) and taking the near horizon limit. This way the complete F $\perp$ D4 
$\perp$ D0 solution takes the form,
\bea\label{fd4d0}
ds^2 &=& H_1^{\half} H_2^{\half}\left[H_3^{-1}\left(-H_1^{-1}H_2^{-1} dt^2 + (dx^6)^2\right) + H_2^{-1} \sum_{i=2}^5
(dx^i)^2 + dr^2 + r^2 d\Omega_3^2\right]\nn
e^{2\phi} &=& \frac{H_1^{\frac{3}{2}}}{H_2^{\half} H_3}, \qquad B_{[2]}\,\,=\,\, H_3^{-1} dt \wedge dx^6\nn
A_{[1]} &=& H_1^{-1} dt, \qquad A_{[5]}\,\,=\,\, H_2^{-1} dt \wedge dx^2 \wedge \ldots \wedge dx^5
\eea
Here F-string is along $x^6$ and D4-brane is along $x^2,\,x^3,\,x^4,\,x^5$. It can be easily checked that
the near horizon limit of \eqref{fd4d0} gives \eqref{m2m5liffull}. One can get more Lifshitz-like solution
with hyperscaling violation by taking T-dualities on the above F $\perp$ D4 $\perp$ D0 solution. So, for
example, by taking T-duality along one of the common transverse directions, say $x^7$, we can generate
F $\perp$ D5 $\perp$ D1 solution given as,
\bea\label{fd5d1}
ds^2 &=& H_1^{\half} H_2^{\half}\left[H_3^{-1}\left(-H_1^{-1}H_2^{-1} dt^2 + (dx^6)^2\right) + H_2^{-1} \sum_{i=2}^5
(dx^i)^2 + H_1^{-1} H_2^{-1} (dx^7)^2 + dr^2 + r^2 d\Omega_2^2\right]\nn
e^{2\phi} &=& \frac{H_1}{H_2 H_3}, \qquad B_{[2]}\,\,=\,\, H_3^{-1} dt \wedge dx^6\nn
A_{[2]} &=& H_1^{-1} dt \wedge dx^7, \qquad A_{[6]}\,\,=\,\, H_2^{-1} dt \wedge dx^2 \wedge \ldots \wedge dx^5 
\wedge dx^7
\eea
Here F-string is along $x^6$, D5-brane is along $x^2,\,\ldots,\,x^5,\,x^7$ and D1-brane is along $x^7$.
The harmonic functions are given as $H_{1,2,3} = 1 + Q_{1,2,3}/r$, where $Q_{1,2,3}$ are the charges associated
with D1-brane, D5-brane and F-string respectively.  In the near horizon limit $r \to 0$
and further defining the coordinate $u^2=r$, we can write the solution \eqref{fd5d1} as,
\bea\label{fd5d1nh}
ds^2 &=& Q_1^{\half} Q_2^{\half}u^2\left[-\frac{u^2}{Q_1Q_2Q_3}dt^2 + \frac{(dx^6)^2}{u^2Q_3} + \frac{\sum_{i=2}^5
(dx^i)^2}{u^2Q_2} + \frac{(dx^7)^2}{Q_1Q_2} + 4\frac{du^2}{u^2} + d\Omega_2^2\right]\nn
e^{2\phi} &=& \frac{Q_1u^2}{Q_2 Q_3}, \qquad B_{[2]}\,\,=\,\, \frac{u^2}{Q_3} dt \wedge dx^6\nn
A_{[2]} &=& \frac{u^2}{Q_1} dt \wedge dx^7, \qquad A_{[6]}\,\,=\,\, \frac{u^2}{Q_2} dt \wedge dx^2 \wedge \ldots \wedge dx^5 
\wedge dx^7
\eea  
It is clear from \eqref{fd5d1nh} that the part of the metric in square bracket is invariant under the scaling
$t \to \lambda^{-1} t$, $x^{1,\,\ldots,\,5} \to \lambda x^{1,\,\ldots,\,5}$, $u \to \lambda u$. However the full metric
is not scale invariant. This tells us that the metric has Lifshitz scaling (with $z=-1$) with hyperscaling violation. 
To obtain the hyperscaling violation exponent we have to compactify the metric on S$^2$ $\times$ R and 
express the resultant metric in the Einstein frame. The compactified metric has the form,
\be\label{fd5d1nhcom}
ds^2 = Q_1^{\frac{1}{5}} Q_2 Q_3^{\frac{2}{5}} u^{\frac{12}{5}}\left[-\frac{u^2}{Q_1Q_2Q_3}dt^2 + \frac{(dx^6)^2}{u^2Q_3} 
+ \frac{\sum_{i=2}^5 (dx^i)^2}{u^2Q_2} + 4\frac{du^2}{u^2}\right]
\ee
Since under the scaling mentioned above this metric changes as $ds \to \lambda^{\frac{6}{5}} ds \equiv 
\lambda^{\frac{\theta}{d}} ds$, where $d$ is the spatial dimension of the boundary theory (which is 5 in this 
case), we have the hyperscaling violation exponent $\theta=6$. Thus F $\perp$ D5 $\perp$ D1 solution in
the near horizon limit gives a Lifshitz-like metric with $z=-1$, $\theta=6$ and $d=5$.    

We can generate more such intersecting solutions from F $\perp$ D4 $\perp$ D0 by applying T-duality
along $x^2$ direction. The resultant solution is F $\perp$ D3 $\perp$ D1 and then by further taking 
T-duality along $x^7$ direction we get F $\perp$ D4 $\perp$ D2 solution. One can easily check that
both these solutions yield Lifshitz-like metric in the near horizon limit. The former one, F $\perp$
D3 $\perp$ D1, gives zero dimensional Lifshitz metric (which can be recast into AdS$_2$ form) in the
near horizon limit very similar
to F $\perp$ D4 $\perp$ D0 and the latter one, F $\perp$ D4 $\perp$ D2 gives five dimensional Lifshitz
metric in the near horizon limit (and dimensional reduction) with $z=-1$ and $\theta=6$ very similar
to F $\perp$ D5 $\perp$ D1 solution. By applying T-duality on F $\perp$ D3 $\perp$ D1 along $x^3$
direction we can generate intersecting solution F $\perp$ D2 $\perp$ D2 and by further applying
T-duality along $x^7$, we can generate F $\perp$ D3 $\perp$ D3 solutions. Again we find that both
these solutions yield Lifshitz-like metric in the near horizon limit. F $\perp$ D2 $\perp$ D2 gives
zero dimensional Lifshitz metric very similar to F $\perp$ D4 $\perp$ D0 and F $\perp$ D3 $\perp$ D3 
gives five dimensional Lifshitz metric with $z=-1$ and $\theta=6$ very similar to F $\perp$ D5 $\perp$
D1. Therefore, we do not give any further details of these solutions. S-dual
of the type IIB solutions can also give more intersecting solutions of the type discussed here, but
they also yield very similar Lifshitz space-time as their original counterpart and therefore we do not
discuss them here.

\subsection{M5 $\perp$ M5 $\perp$ M5 solution}

Apart from M2 $\perp$ M5 solution the other solution which yields AdS geometry in the near horizon limit
is M5 $\perp$ M5 $\perp$ M5. The solution has the form \cite{Klebanov:1996mh},
\bea\label{m5m5m5}
ds^2 &=& (H_2H_3H_4)^{\frac{2}{3}}\left[(H_2H_3H_4)^{-1}(-dt^2 + (dx^1)^2) + (H_2H_3)^{-1}((dx^2)^2+(dx^3)^2)\right.\nn
& & \left. + (H_2H_4)^{-1}((dx^4)^2 + (dx^5)^2)
+ (H_3H_4)^{-1}((dx^6)^2 + (dx^7)^2) + dr^2 + r^2 d\Omega_2^2\right]\nn
A_{[6]} &=& H_2^{-1} dt \wedge dx^1 \wedge \ldots \wedge dx^5, \qquad A_{[6]}' \,\,=\,\, H_3^{-1} dt 
\wedge dx^1 \wedge dx^4 \wedge \ldots \wedge dx^7\nn
A_{[6]}'' &=& H_4^{-1} dt \wedge dx^1 \wedge dx^2 \wedge dx^3 \wedge dx^6 \wedge dx^7
\eea
Note that here the three M5-branes intersect on a string along $x^1$ and they intersect pairwise
on 3-branes along $x^1,\,x^2,\,x^3$, along $x^1,\,x^4,\,x^5$ and along $x^1,\,x^6,\,x^7$. The three harmonic functions
are given as $H_{2,3,4} = 1 + Q_{2,3,4}/r$, where $Q_{2,3,4}$ are the charges associated with the
three M5-branes. The M5-branes are electric and so, $A_{[6]}$, $A_{[6]}'$ and
$A_{[6]}''$ are the three 6-form gauge fields to which they couple. In the near horizon limit
$H_{2,3,4} \approx Q_{2,3,4}/r$ and further defining a new coordinate by $u^2 = 1/r$, we can rewrite
the metric in \eqref{m5m5m5} as
\bea\label{m5m5m5nh}
ds^2 &=& (Q_2Q_3Q_4)^{\frac{2}{3}}\left[\frac{-dt^2 + (dx^1)^2 + 4Q_2Q_3Q_4 du^2}{Q_2Q_3Q_4 u^2} + 
\frac{(dx^2)^2+(dx^3)^2}{Q_2Q_3}\right.\nn
& & \left. + \frac{(dx^4)^2 + (dx^5)^2}{Q_2Q_4}
+ \frac{(dx^6)^2 + (dx^7)^2}{Q_3Q_4} + d\Omega_2^2\right]
\eea
It is clear from the metric \eqref{m5m5m5nh} that it has the structure AdS$_3$ $\times$ E$^6$ $\times$
S$^2$. As discussed in section 2, we will show how starting from this AdS geometry we can generate
Schr\" odinger/Lifshitz dual space-times by solution generating techniques.

To obtain Schr\" odinger space-times we generate pp-waves along $x^1$ by standard technique. The above
metric in that case takes the form,
\bea\label{m5m5m5nhwave}
ds^2 &=& (Q_2Q_3Q_4)^{\frac{2}{3}}\left[\frac{-dt^2 + (dx^1)^2 +(H_1-1)(dt-dx^1)^2 + 
4Q_2Q_3Q_4 du^2}{Q_2Q_3Q_4 u^2}\right.\nn 
& & \left. +
\frac{(dx^2)^2+(dx^3)^2}{Q_2Q_3}
+ \frac{(dx^4)^2 + (dx^5)^2}{Q_2Q_4}
+ \frac{(dx^6)^2 + (dx^7)^2}{Q_3Q_4} + d\Omega_2^2\right]
\eea 
where $H_1=1+Q_1/r$ is another harmonic function and $Q_1$ is the asymptotic charge
carried by the wave. In terms of $u$, harmonic function is given as $H_1=1+Q_1u^2$.
Substituting this in the metric \eqref{m5m5m5nhwave}, going to the light cone coordinates
defined earlier and further taking the double Wick rotation ($t \to i\xi$ and $\xi \to -it$),
the metric as well as the gauge fields take the forms,
\bea\label{m5m5m5nhwavesch}
ds^2 &=& (Q_2Q_3Q_4)^{\frac{2}{3}}\left[-\frac{2Q_1}{Q_2Q_3Q_4} dt^2 + \frac{-2 d\xi dt+
4Q_2Q_3Q_4 du^2}{Q_2Q_3Q_4 u^2}\right.\nn
& & \left. +
\frac{(dx^2)^2+(dx^3)^2}{Q_2Q_3}
+ \frac{(dx^4)^2 + (dx^5)^2}{Q_2Q_4}
+ \frac{(dx^6)^2 + (dx^7)^2}{Q_3Q_4} + d\Omega_2^2\right]\nn
A_{[6]} &=& \frac{1}{Q_2u^2}dt \wedge d\xi \wedge \ldots \wedge dx^5, \qquad A_{[6]}' \,\,=\,\, 
\frac{1}{Q_3u^2} dt
\wedge d\xi \wedge dx^4 \wedge \ldots \wedge dx^7\nn
A_{[6]}'' &=& \frac{1}{Q_4u^2} dt \wedge d\xi \wedge dx^2 \wedge dx^3 \wedge dx^6 \wedge dx^7
\eea 
Under the scaling symmetry $t \to \lambda^0 t$, $\xi \to \lambda^2 \xi$ and $u \to \lambda u$,
the metric in \eqref{m5m5m5nhwavesch} remains invariant. Note that the other coordinates do not
scale and therefore we can compactify the metric on E$^6$ $\times$ S$^2$, to get a zero dimensional
Schr\" odinger metric with $z=0$ and $\theta=0$, very much like the case we discussed for M2 $\perp$
M5 solution in subsection 4.1.

To obtain Lifshitz space-time we rewrite the metric \eqref{m5m5m5nhwave} as follows,
\bea\label{m5m5m5nhwave1}
ds^2 &=& (Q_2Q_3Q_4)^{\frac{2}{3}}\left[\frac{-H_1^{-1}dt^2 +H_1\left((1-H_1^{-1})dt-dx^1\right)^2 + 
4Q_2Q_3Q_4 du^2}{Q_2Q_3Q_4 u^2}\right.\nn 
& & \left. +
\frac{(dx^2)^2+(dx^3)^2}{Q_2Q_3}
+ \frac{(dx^4)^2 + (dx^5)^2}{Q_2Q_4}
+ \frac{(dx^6)^2 + (dx^7)^2}{Q_3Q_4} + d\Omega_2^2\right]
\eea 
Compactifying along $x^1$, the wave direction, we obtain the string theory solution as,
\bea\label{m5m5m5nhwavelif}
ds^2 &=& (Q_1Q_2Q_3Q_4)^{\half}\left[-\frac{dt^2}{Q_1Q_2Q_3Q_4 u^4} + 
\frac{4du^2}{u^2} +
\frac{(dx^2)^2+(dx^3)^2}{Q_2Q_3}\right.\nn
& & \left. + \frac{(dx^4)^2 + (dx^5)^2}{Q_2Q_4}
+ \frac{(dx^6)^2 + (dx^7)^2}{Q_3Q_4} + d\Omega_2^2\right]\nn
e^{2\phi} &=& \frac{Q_1^{\frac{3}{2}}}{(Q_2Q_3Q_4)^{\half}}\nn
A_{[5]} &=& \frac{1}{Q_2u^2}dt \wedge dx^2 \wedge \ldots \wedge dx^5, \qquad A_{[5]}' \,\,=\,\, 
\frac{1}{Q_3u^2} dt
\wedge dx^4 \wedge \ldots \wedge dx^7\nn
A_{[5]}'' &=& \frac{1}{Q_4u^2} dt \wedge dx^2 \wedge dx^3 \wedge dx^6 \wedge dx^7, \qquad A_{[1]} \,\,=\,\,
\frac{1}{Q_1u^2} dt
\eea
The metric above has a scaling symmetry $t \to \lambda^2 t$, $u \to \lambda u$. Since the other spatial
coordinates do not scale we can compactify them to obtain a zero dimensional Lifshitz metric. However, we
note that by defining a new parameter $\tilde u = u^2$ we can recast the metric into AdS$_2$ form. The dilaton
in this case is constant very much like M2 $\perp$ M5 case we discussed before. By looking at the solution
\eqref{m5m5m5nhwavelif} we recognize this to be the near horizon limit of the intersecting 1/16 BPS type
IIA string theory solution D4 $\perp$ D4 $\perp$ D4 $\perp$ D0. By applying T- and S-dualities to this solution
one can construct many such intersecting 1/16 BPS solutions. These solutions in the near horizon limit will
not give AdS geometry, but it will be worthwhile to see whether some of these solutions can give rise to
interesting Lifshitz-like space-time. We leave this for a future investigation.
    
\subsection{D1 $\perp$ D5 solution}

This type IIB string theory solution is known to lead to AdS$_3$ $\times$ E$^4$ $\times$ S$^3$ metric in the
near horizon limit. So, we can obtain Schr\" odinger/Lifshitz dual space-times starting from this metric. In
order to see this we first write the solution,
\bea\label{d1d5}
ds^2 &=& H_2^{\half}H_3^{\half}\left[H_2^{-1}H_3^{-1} (-dt^2 + (dx^1)^2) + H_3^{-1}\sum_{i=2}^5 (dx^i)^2 + dr^2 + 
r^2 d\Omega_3^2\right]\nn
e^{2\phi} &=& \frac{H_2}{H_3}\nn
A_{[2]} &=& H_2^{-1} dt \wedge dx^1, \qquad A_{[6]} \,\,=\,\, H_3^{-1} dt \wedge dx^1 \wedge \ldots \wedge dx^5
\eea
Here $H_{2,3} = 1 + Q_{2,3}/r^2$ and $Q_{2,3}$ are the charges associated with D1-brane and D5-brane respectively.
D1-brane lies along $x^1$ and D5-brane lies along $x^1,\,\ldots,\,x^5$. In the near horizon limit $H_{2,3} \approx
Q_{2,3}/r^2$ and then defining a new coordinate $u = 1/r$, we write the solution as,
\bea\label{d1d5nh}
ds^2 &=& Q_2^{\half}Q_3^{\half}\left[\frac{-dt^2 + (dx^1)^2 + Q_2Q_3 du^2}{Q_2Q_3u^2} + \frac{1}{Q_3}\sum_{i=2}^5 (dx^i)^2 + d\Omega_3^2\right]\nn
e^{2\phi} &=& \frac{Q_2}{Q_3}\nn
A_{[2]} &=& \frac{1}{Q_2u^2} dt \wedge dx^1, \qquad A_{[6]} \,\,=\,\, \frac{1}{Q_3u^2} dt \wedge dx^1 \wedge \ldots \wedge dx^5
\eea     
The metric above is AdS$_3$ $\times$ E$^4$ $\times$ S$^3$. Once we have AdS$_3$, we can get Schr\" odinger space-time as 
before by first generating a pp-wave along the common brane direction $x^1$, going to the light-cone coordinates defined
before and then taking a double Wick rotation. The resultant Schr\" odinger metric takes the form,
\be\label{d1d5sch}
ds^2 = Q_2^{\half}Q_3^{\half}\left[-\frac{2Q_1}{Q_2Q_3}dt^2 + \frac{-2d\xi dt + Q_2Q_3 du^2}{Q_2Q_3u^2} + \frac{1}{Q_3}\sum_{i=2}^5 (dx^i)^2 
+ d\Omega_3^2\right]
\ee
where $Q_1$ is the asymptotic momentum carried by the wave.
The metric is invariant under the scaling $t \to \lambda^0 t$, $\xi \to \lambda^2 \xi$ and $u \to \lambda u$. Also, since the other
spatial coordinates do not scale we can compactify the solution on T$^4$ $\times$ S$^3$ to obtain a zero dimensional Schr\" odinger
space-time very similar to M2 $\perp$ M5 case.

Lifshitz space-time can also be obtained as before by rewriting the metric with waves in a suitable form as described in section 2
and then taking T-duality along the wave direction $x^1$. The resultant metric will be the near horizon limit of the T-dual 
solution of `D1 $\perp$ D5 + wave' solution we just described. This T-dual solution is nothing but the F $\perp$ D4 $\perp$ D0 solution
\eqref{fd4d0} we described in subsection 4.2. In the near horizon limit it gives a zero dimensional Lifshitz metric which can be written
in AdS$_2$ form by a suitable coordinate transformation as shown in \eqref{m2m5liffull}. 

\subsection{D2 $\perp$ D4 solution}

This is a type IIA string theory solution which can be obtained from D1 $\perp$ D5 solution by applying T-duality
along one of the D5-brane direction transverse to D1-brane. Here D2 and D4 intersect on a string and is 1/4 BPS,
unlike 1/2 BPS D2-D4 solution where D2-brane is completely inside the D4-brane. The solution has been discussed
in \cite{Dey:2012rs}, where we found that in the near horizon limit it gives AdS$_3$ $\times$ E$^4$ $\times$ S$^3$ as in D1
$\perp$ D5 system. We can generate Schr\" odinger/Lifshitz dual space-times starting from this AdS$_3$ geometry.
As in D1 $\perp$ D5 case here also we get zero dimensional Schr\" odinger metric with dynamical critical exponent $z=0$ 
and with no hyperscaling violation. The corresponding Lifshitz is also zero dimensional with $z=2$ and $\theta=0$, which
can also be recast into AdS$_2$ form.  

D3 $\perp$ D3 solution of type IIB string theory \cite{Dey:2012rs} is also known to give AdS$_3$ in the near horizon limit.
The Schr\" odinger/Lifshitz dual space-times obtained from this solution also have very similar forms as those of
D2 $\perp$ D4 or D1 $\perp$ D5 solutions. So, we do not give any further details for this solution.

\subsection{F $\perp$ NS5 solution: an exception}

F $\perp$ NS5 solution can be obtained from D1 $\perp$ D5 solution by applying S-duality. Like D1 $\perp$ D5,
this solution also leads to AdS metric in the near horizon limit. F $\perp$ NS5 solution of either type IIA or
IIB string theory has the form,
\bea\label{fns5}
ds^2 &=& H_3\left[H_2^{-1}H_3^{-1} (-dt^2 + (dx^1)^2) + H_3^{-1}\sum_{i=2}^5 (dx^i)^2 + dr^2 + 
r^2 d\Omega_3^2\right]\nn
e^{2\phi} &=& \frac{H_3}{H_2}\nn
B_{[2]} &=& H_2^{-1} dt \wedge dx^1, \qquad B_{[6]} \,\,=\,\, Q_3\,{\rm Vol}(\Omega_3)
\eea
where $H_{2,3} = 1 + Q_{2,3}/r^2$ and $Q_{2,3}$ are the charges associated with the F-string and the NS5-brane
respectively. In the near horizon limit $H_{2,3} \approx Q_{2,3}/r^2$ and then defining a new coordinate by
$u = 1/r$ we can write the metric in \eqref{fns5} as,
\be\label{fns5nh}
ds^2 = Q_3\left[\frac{-dt^2 + (dx^1)^2 + Q_2Q_3 du^2}{Q_2Q_3u^2} + \frac{1}{Q_3}\sum_{i=2}^5 (dx^i)^2 +  
d\Omega_3^2\right] 
\ee
It is clear that the metric \eqref{fns5nh} has AdS$_3$ $\times$ E$^4$ $\times$ S$^3$ structure. As we argued
in section 2, we might expect to get Schr\" odinger/Lifshitz dual space-times starting from this AdS geometry.
However, we will argue that this is not the case and F $\perp$ NS5 is an exception. Schr\" odinger space-time can
be obtained by generating a wave along the common brane direction $x^1$, then rewriting the resulting solution in
the light-cone coordinates and finally taking the double Wick rotation. As in D1 $\perp$ D5 case, here also we
get a zero dimensional Schr\" odinger metric with $z=0$ and $\theta=0$. We can then try to get Lifshitz space-time
by rewriting the metric as given in \eqref{prelif} and then taking T-duality along the wave direction $x^1$. However,
since the solution `F $\perp$ NS5 + wave' is invariant under T-duality, we can not generate Lifshitz metric in this
case and the resultant metric will still exhibit Schr\" odinger symmetry. So, this is the only case where our algorithm
of generating Schr\" odinger/Lifshitz dual space-times starting from AdS geometry does not work.

\section{Conclusion}

To conclude, in this paper we have shown how starting from AdS geometry (obtained from various string/M theory
solutions in the near horizon limit either directly or upto a conformal transformation), one can generate 
Schr\" odinger/Lifshitz dual space times (without or with
hyperscaling violation) by some solution generating techniques known for string/M theory. Schr\" odinger/Lifshitz 
space-times obtained in this way are dual in the sense that their dynamical critical exponents add upto 2.
We have studied various examples including simple branes and double and triple intersecting branes. Among simple
brane solutions we have studied M2-, M5- and D3-branes, which are known to give AdS$_4$, AdS$_7$ and AdS$_5$ geometries 
respectively, directly in the near horizon
limit. We have also studied D$(p+1)$-branes ($p\neq 3,4$) which are known to give AdS$_{p+3}$ geometry upto a conformal
transformation. Then we have studied double and triple intersecting M-brane solutions M2 $\perp$ M5 and M5 $\perp$ M5
$\perp$ M5 which are known to give AdS$_3$ geometry and some double intersecting string solutions D1 $\perp$ D5 (of type IIB),
D2 $\perp$ D4 (of type IIA) and F $\perp$ NS5 (of type IIA or IIB) which are also known to give AdS$_3$ geometries.
In all these cases we have obtained Schr\" odinger/Lifshitz dual space-times except F $\perp$ NS5 case.
For obtaining Schr\" odinger space-times we had to deform the AdS geometry by introducing pp-waves along one of the brane
directions (for single brane) and one of the common brane directions (for the intersecting branes) and then wrote the
solution in the light-cone coordinate. By further taking a double Wick rotation we obtained metric having Schr\" odinger
symmetry without or with hyperscaling violation. On the other hand to obtain Lifshitz space-times we took the deformed
solution in Poincare coordinates and either dimensionally reduced it along the wave direction (for M-theory solutions) or
took T-duality (for string theory solutions) along the same direction. These reduced or T-dual solutions then exhibited 
Lifshitz scaling symmetry without or with hyperscaling violation. More Lifshitz (but not Schr\" odinger) space-times were 
obtained by taking further T-duality along other directions as shown for specific case in subsection 4.2. 
Thus starting from the same string or M-theory solutions, which are known to give AdS geometry
in the near horizon decoupling limit, we obtained both Schr\" odinger and Lifshitz space-times by using solution 
generating techniques. Since these solution generating techniques do not break supersymmetry and the string/M-theory solutions
are BPS, we expect the Schr\" odinger and the Lifshitz space-times we obtained this way also preserve some fraction of space-time 
supersymmetries. Intersecting branes give low dimensional or AdS$_3$ space and consequently the
Schr\" odinger/Lifshitz dual space-times obtained in these cases are zero dimensional. Zero dimensional Lifshitz can also be cast    
into AdS$_2$ form by some coordinate transformation. However, we found that we can not obtain Schr\" odinger/Lifshitz dual
space-times for F $\perp$ NS5 intersecting solution even if it gives AdS$_3$ geometry (in the near horizon limit). We can get 
Schr\" odinger metric, but no Lifshitz metric can be obtained by T-duality as this solution with pp-wave (along the common 
brane direction) is T-duality invariant. Our results can be interpreted from the dual
field theory point of view as obtaining certain strongly correlated condensed matter system (having Schr\" odinger 
or Lifshitz scaling symmetry) by some sort of a deformation of a relativistic system. The precise form of the 
deformation in field theory is not clear to us.     

\vspace{.2cm}

\noindent{\large \bf Note added:}

\vspace{.2cm}

After submitting this paper to the archive, we were informed by the authors of \cite{Gath:2012pg} that
there is quite a bit of overlap of this paper with sections 5 and 6 of \cite{Gath:2012pg}. In section 2
we have shown that the dynamical critical exponents of Schr\" odinger/Lifshitz dual space-times add upto
2 and this was also observed in section 5 of their paper. Also it can be checked that their general formula
of $z$ and $\theta$ given in eqs.(6.43) and (6.44) for Lifshitz space-times in \cite{Gath:2012pg} match with 
the results given in sections 3 and 4 of our paper for different values of the parameters defined in 
\cite{Gath:2012pg}. We are grateful to Jakob Gath, Jelle Hartong, Ricardo Monteiro and Niels Obers for 
informing us about this.   

\section*{Acknowledgements}

One of the authors (PD) would like to acknowledge thankfully the financial
support of the Council of Scientific and Industrial Research, India
(SPM-07/489 (0089)/2010-EMR-I). We would like to thank Harvendra Singh for useful
discussions. We would also like to thank K. Narayan for the suggestions about rearranging
some of the references in an earlier version of this paper and Domenico Orlando for informing
us about \cite{Israel:2004vv}.

\end{document}